\documentclass[pdflatex,sn-nature]{sn-jnl}

\usepackage{graphicx}%
\usepackage{multirow}%
\usepackage{amsmath,amssymb,amsfonts}%
\usepackage{amsthm}%
\usepackage{mathrsfs}%
\usepackage[title]{appendix}%
\usepackage{xcolor}%
\usepackage{textcomp}%
\usepackage{manyfoot}%
\usepackage{booktabs}%
\usepackage{algorithm}%
\usepackage{algorithmicx}%
\usepackage{algpseudocode}%
\usepackage{listings}%
\usepackage{dsfont}
\usepackage{anyfontsize}
\usepackage{subcaption}
\usepackage{makecell}

\theoremstyle{thmstyleone}%
\theoremstyle{thmstyletwo}%

\theoremstyle{thmstylethree}%

\begin{document}

\title[Article Title]{Uncertainty-Aware Multimodal Deep Learning for Prediction of Radiation-Induced Xerostomia in Head and Neck Cancer}

\author[1]{\fnm{Jin} \sur{Yang, PhD}}

\author[1]{\fnm{Tian} \sur{Liu, PhD}}

\author[2]{\fnm{Jos\'{e} A.} \sur{Pe\~{n}agar\'{i}cano, MD}}

\author[1]{\fnm{Jing} \sur{Wang, PhD}}

\author[3]{\fnm{Erin L.} \sur{Moshier, MSc}}

\author[1]{\fnm{Jerry} \sur{Liu, MD}}

\author[1]{\fnm{Richard} \sur{Bakst, MD}}

\author*[1]{\fnm{Ming} \sur{Chao, PhD}}\email{ming.chao@mountsinai.org}

\affil[1]{\orgdiv{Department of Radiation Oncology}, \orgname{Icahn School of Medicine at Mount Sinai}, \orgaddress{\city{New York}, \state{NY} \postcode{10029}, \country{USA}}}

\affil[2]{\orgdiv{Department of Radiation Oncology}, \orgname{H. Lee Moffitt Cancer Center \& Research Institute}, \orgaddress{\city{Tampa}, \state{FL} \postcode{33612}, \country{USA}}}

\affil[3]{\orgdiv{Department of Population Health Science and Policy}, \orgname{Icahn School of Medicine at Mount Sinai}, \orgaddress{\city{New York}, \state{NY} \postcode{10029}, \country{USA}}}

\abstract{
\textbf{Purpose:} To develop and evaluate an uncertainty-aware multimodal learning framework for predicting radiation-induced xerostomia in patients receiving head-and-neck cancer radiotherapy.

\textbf{Methods and Materials:} We develop a Bidirectional Attention-enhanced Diffusion Transformer (BA-DiT) that integrates self-supervised multimodal representation learning with supervised Bayesian fine-tuning for individualized xerostomia risk prediction. The model primarily focuses on bilateral parotid glands while incorporating surrounding anatomical and dosimetric information from pretreatment CT images and corresponding three-dimensional dose distributions. Specifically, bidirectional cross-attention integrates anatomical and dosimetric features which are extracted by encoders with an identical architecture, and subsequently a latent Diffusion Transformer learns joint representations. These components are optimized during pretraining through diffusion-based learning, modality-specific feature reconstruction, and contrastive alignment. During supervised fine-tuning, a Bayesian classifier estimates individualized xerostomia risk and quantifies predictive uncertainty through approximate Bayesian inference. A learnable label noise transition layer is additionally incorporated to measure potential uncertainty in observed toxicity labels. BA-DiT was evaluated using four independent cohorts from multiple institutions (Moffitt, Emory, MSHS, and TCIA) and evaluation included predictive performance, probability calibration, uncertainty quantification, decision curve analysis, and representative patient-level case analyses.

\textbf{Results:} In independent MSHS and Emory cohorts, BA-DiT achieved the highest areas under the receiver operating characteristic curve (AUCs) of $0.83$ and $0.80$, respectively, and outperformed baseline models in accuracy, sensitivity, and F1 score. Uncertainty quantification revealed that total predictive uncertainty was highest near the decision boundary, with aleatoric uncertainty constituting the predominant component across predicted risk levels. Furthermore, uncertainty-based selective prediction improved the net benefit of BA-DiT among patients retained for prediction, with greater improvements observed at lower prediction coverage. 

\textbf{Conclusions:} BA-DiT enables individualized prediction of radiation-induced xerostomia while providing complementary estimates of predictive uncertainty. By identifying predictions associated with greater uncertainty, this framework may support more informed clinical decision-making and facilitate personalized strategies for xerostomia risk management.
}




\maketitle

\section{Introduction}\label{sec1}
Xerostomia remains a debilitating toxicity of radiation therapy (RT) for head-and-neck cancer (HNC), affecting up to $80\%$ of patients and impairing oral health, speech, swallowing, and overall quality of life \cite{wijers2002patients,dirix2006radiation, pow2006xerostomia, vissink2010clinical, liu2011risk, nathan2023current}, despite advances in intensity-modulated RT (IMRT) and proton therapy \cite{eisbruch1999dose, romesser2016proton, sherry2021proton, cao2026intensity}. Current toxicity mitigation relies primarily on established salivary-gland dose constraints, including a mean parotid dose $<$ 26Gy recommended by the Quantitative Analysis of Normal Tissue Effects in the Clinic (QUANTEC) \cite{deasy2010radiotherapy, beetz2012ntcp, beetz2014quantec, katsoulakis2022consensus, van2023prediction}. Although these constraints have improved salivary-gland sparing, they are largely population-derived and cannot fully account for individual differences in gland susceptibility and functional response. Consequently, substantial interpatient variability in post-RT salivary dysfunction persists, even among patients receiving comparable gland doses \cite{beetz2012ntcp,bussels2004doseresponse}. 

Adaptive radiotherapy (ART) seeks to optimize treatment delivery in response to intratreatment anatomical changes, such as weight loss, tumor regression, and organ-at-risk (OAR) deformation, through updated target and OAR contours and treatment-plan recalculation \cite{chao2010image, raghavan2016anatomic, lalonde2021anatomic, rios2026quantification}. However, the phase III ARTIX trial demonstrated that weekly anatomy-driven ART did not significantly reduce xerostomia compared with standard IMRT \cite{castelli2023weekly}. This finding highlights the need to complement treatment-plan adaptation with approaches that better characterize individual susceptibility to treatment-related toxicity.

Multimodal artificial intelligence (AI) offers an opportunity to integrate complementary clinical, imaging, and dosimetric information for individualized prediction of RT-induced toxicity. Existing approaches include radiomics and conventional machine-learning models based on handcrafted imaging and dosimetric features \cite{rosen2018early,chao2022cluster, berger2022predicting, chao2025exploring, xue2026advRO}, as well as deep learning models that learn predictive representations directly from multimodal data \cite{men2019deep, Chu2024IJROBP, wan2026deep}. Although these approaches have shown promise in xerostomia risk stratification, predictive performance alone does not establish how reliably an individual risk estimate can be interpreted. Most AI models provide point estimates or risk probabilities without explicitly quantifying predictive uncertainty, leaving clinicians with limited information about the confidence that should be placed in individual predictions. This limitation is particularly relevant in heterogeneous clinical populations, where patient characteristics, treatment response, and available data may differ from those represented during model development. 

Predictive uncertainty provides important information for interpreting individual risk estimates and identifying predictions that may warrant additional clinical review \cite{wahid2024artificial,dunger2025uncertainties}. It can arise from inherent variability in clinical observations (aleatoric uncertainty), limited model knowledge due to finite training data (epistemic uncertainty), ambiguous toxicity labels, and distributional differences between model development and deployment populations \cite{kendall2017uncertainties,nair2020exploring,lambert2024trustworthy}. Although various approaches to uncertainty quantification have been developed in AI, including ensemble-based methods, test-time augmentation, and Bayesian methods \cite{dunger2025uncertainties, lakshminarayanan2017simple, wang201934, blundell2015weight}, their application to treatment outcome prediction, particularly RT-induced toxicity modeling, remains limited. Bayesian learning offers a principled framework for characterizing uncertainty in model parameters, while Monte Carlo sampling enables estimation of predictive variability across stochastic forward passes \cite{gal2016dropout}. Combining multimodal deep learning with Bayesian uncertainty quantification may therefore yield individualized toxicity risk estimates together with information about the confidence with which those predictions should be interpreted, supporting more informed and risk-aware clinical decision-making. However, this integration is largely unexplored for individualized xerostomia prediction.

To address this unmet clinical need, we developed a Bidirectional Attention-enhanced Diffusion Transformer (BA-DiT) framework for uncertainty-aware xerostomia prediction. Using pretreatment CT images and corresponding planning three-dimensional (3D) radiation dose distributions, BA-DiT learns spatially informed representations of the parotid glands, a major determinant of post-RT salivary dysfunction. The framework integrates self-supervised representation learning to leverage heterogeneous multi-institutional imaging and dosimetric data while reducing reliance on labeled toxicity outcomes, followed by supervised Bayesian fine-tuning to estimate individualized xerostomia risk and associated predictive uncertainty.

We evaluated BA-DiT using four independent patient cohorts from multiple institutions to assess its predictive performance, uncertainty estimation, and generalizability across heterogeneous clinical datasets. By leveraging routinely available pretreatment information, this study investigates the potential of uncertainty-aware AI to identify patients at increased risk of xerostomia before treatment initiation, with the ultimate goal of supporting more informed and personalized RT planning for patients with HNC.

\section{Methods and Materials}
\subsection{Patient Cohorts and Data Pre-processing}
The study comprised patients with HNC who received RT using fixed-beam IMRT or volumetric modulated arc therapy (VMAT). Primary tumor sites spanned the larynx, hypopharynx, oropharynx, nasopharynx, oral cavity/tongue, nasal cavity, and salivary glands. Prescribed treatment regimens consisted of either 70Gy delivered to the planning target volume (PTV) in 35 fractions of 2Gy or 66Gy delivered in either 33 fractions of 2Gy or 30 fractions of 2.2Gy. 

Xerostomia was assessed at 6 months after RT using the Common Terminology Criteria for Adverse Events (CTCAE), version 4.0, based on physician documentation of patient-reported dry-mouth symptoms during routine follow-up visits. Although no formal interobserver or outcome-adjudication quality assurance was performed, experienced HNC specialists assessed and documented toxicity according to standard clinical protocols. Patients who were lost to follow-up or unable to provide outcome information at the 6-month assessment were excluded. Baseline xerostomia grades were also recorded to distinguish pre-existing symptoms from post-RT toxicity. Patients were classified into four severity grades: grade 0 (no dry-mouth), grade 1 (mild), grade 2 (moderate), and grade 3 (severe). For model development and evaluation, a binary clinical endpoint was defined as no or mild xerostomia (grade $\leq$ 1) versus moderate-to-severe xerostomia (grade $\geq$ 2). Each patient record included a pretreatment CT image set, the corresponding 3D dose distribution, and segmentation masks of the left and right parotid glands. 

The Moffitt Cancer Center cohort comprised 156 patients, of whom 77 (49.4\%) experienced no or mild xerostomia (grade $\leq 1$) and 79 (50.6\%) developed moderate-to-severe xerostomia (grade $\geq 2$). The Emory University cohort included 147 patients, with 50 (34.0\%) classified as having grade $\leq 1$ xerostomia and 97 (66.0\%) having grade $\geq 2$ xerostomia. The Mount Sinai Health System (MSHS) cohort comprised 63 patients, including 25 (39.7\%) with grade $\leq 1$ and 38 (60.3\%) with grade $\geq 2$ xerostomia. The final cohort was obtained from The Cancer Imaging Archive (TCIA) and included 27 patients \cite{grossberg2020hnscc}. Of these, 22 (81.5\%) had grade $\leq 1$ and 5 (18.5\%) had grade $\geq 2$ xerostomia.

Prior to model training, the CT images and dose distributions underwent a multistep preprocessing pipeline. First, CT voxel intensities were clipped to a window of $(-200, 300)$ HU to reduce the influence of outliers and noise while emphasizing the parotid regions. Similarly, dose values were clipped to the range of $(0, 1.5)$, normalized to a reference dose of $70$ Gy, to suppress noise and exclude implausible values. Following intensity clipping, both modalities were cropped to the head-and-neck region using mask-guided cropping based on the left and right parotid segmentation masks. The resulting volumes were then resampled to a fixed spatial dimension of $64\times160\times160$ voxels. Finally, voxel values were normalized via z-score normalization to promote numerical stability and facilitate robust feature learning. 

To improve model generalizability and reduce overfitting, data augmentation was applied during training. Specifically, CT and dose patches were mirrored along the  $x$- and $y$-axes, each with a probability of $0.5$, and subjected to affine transformations with a probability of 0.3. In addition, CT patches were independently augmented with zero-centered additive Gaussian noise, with variance sampled from a uniform distribution $U(0, 0.01)$, and brightness adjustments, applied with probabilities of $0.2$ and $0.3$, respectively.

\begin{figure*}[!ht]
	\centering
	\includegraphics[width=\linewidth]{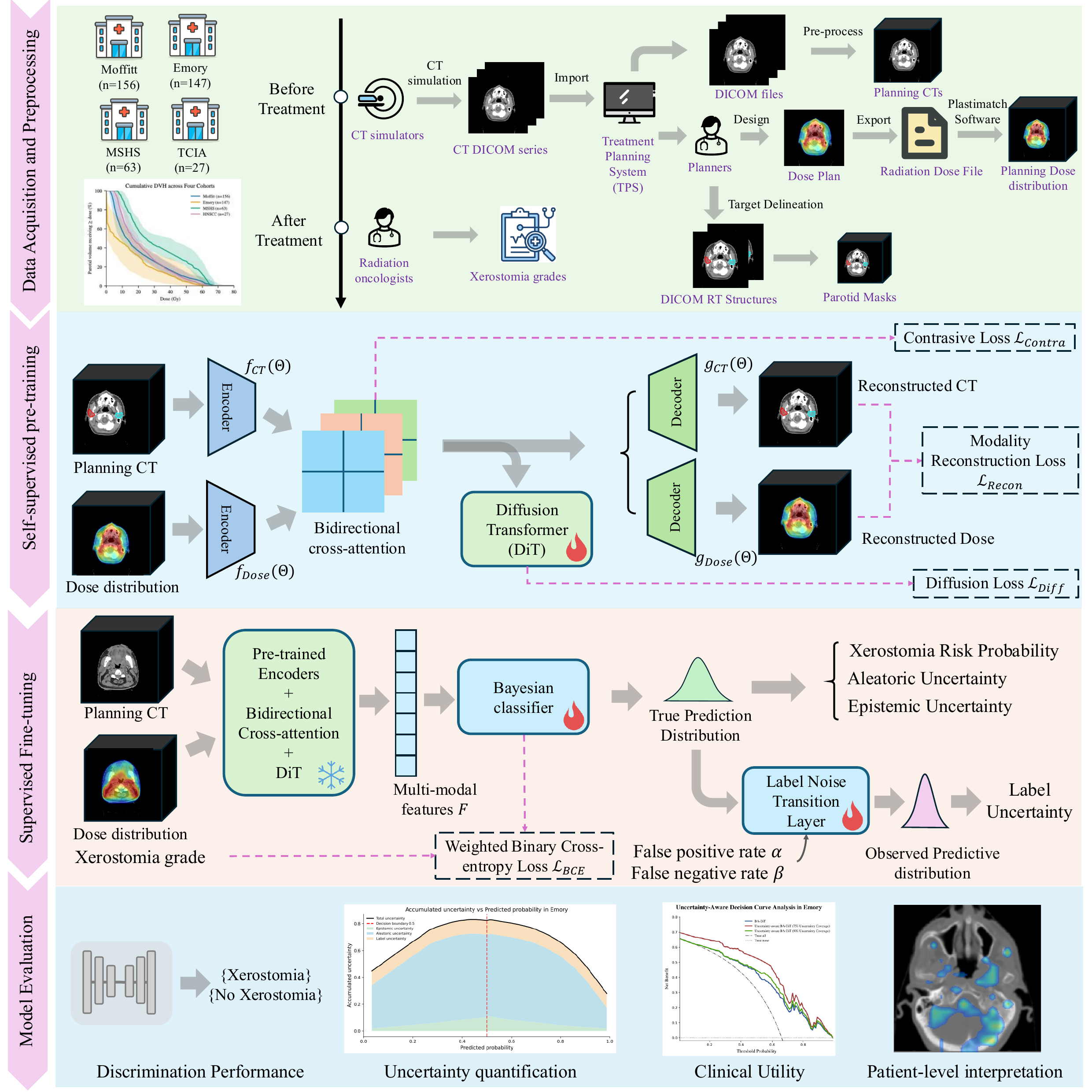} \\
	\caption{\textbf{Overview of the Bidirectional Attention-enhanced Diffusion Transformer framework (BA-DiT) for xerostomia risk prediction and uncertainty quantification}. The framework consists of four modules: (1) data acquisition and preprocessing; (2) self-supervised pretraining; (3) supervised fine-tuning; and (4) model evaluation. Further technical details are provided in the Methods.}
	\label{fig:model}
\end{figure*}

\subsection{Model Implementation}
BA-DiT is a multimodal deep learning framework comprising four module: (1) data acquisition and preprocessing; (2) self-supervised pretraining; (3) supervised fine-tuning; and (4) model evaluation (Figure \ref{fig:model}). (1) Data acquisition and preprocessing: Pretreatment CT images, corresponding planning dose distributions, parotid gland segmentation masks, and xerostomia outcomes were collected from four cohorts. CT and dose files were processed using Plastimatch to generate the model inputs. (2) Self-supervised pretraining: CT images and dose distributions are independently processed by modality-specific encoders, and the resulting features are integrated through bidirectional cross-attention. The fused multimodal features are subsequently processed by a latent Diffusion Transformer (DiT) to learn spatially informed representations of CT anatomy and spatial dose distributions across the multi-institutional datasets. These components are optimized during self-supervised pretraining via contrastive learning, modality-specific reconstruction, and diffusion-based learning. (3) Supervised fine-tuning: The pretrained encoders, bidirectional cross-attention, and DiT are integrated with a Bayesian classifier for xerostomia risk prediction and quantification of associated aleatoric and epistemic uncertainties. A learnable label noise transition layer is additionally employed to model label uncertainty. (4) Model evaluation: BA-DiT was evaluated in terms of discrimination performance, uncertainty quantification, clinical utility, and patient-level interpretation. The primary technical details are described in the following sections.

\textbf{Bidirectional Attention-enhanced Diffusion Transformer.} BA-DiT takes pretreatment CT images and 3D dose distributions as inputs and employs two convolutional neural network (CNN)-based encoders with the identical architecture to extract modality-specific features. Each encoder consists of three sequential downsampling blocks, each comprising a 3D convolutional layer with a kernel size of $3\times3\times3$ and a stride of $2$, followed by batch normalization and ReLU activation. The number of feature channels increases progressively from 32 to 64 and 128, with the spatial dimensions reduced by a factor of 2 at each block. An additional convolutional layer with a kernel size of $3\times3\times3$ and a stride of 1, followed by instance normalization and ReLU activation, maps the resulting feature maps to 256 channels.

The encoded CT and dose features are integrated subsequently using a bidirectional cross-attention fusion module. First, adaptive average pooling is independently applied to each modality to obtain a $4\times4\times4$ spatial grid. The pooled features are flattened into sequences of 64 tokens per modality and linearly projected into a shared 256-dimensional latent embedding space. A shared learnable spatial positional embedding is then added to both tokens sequences to preserve spatial correspondence. Bidirectional cross-attention fusion is implemented through two parallel attention streams. In the first stream, CT tokens serve as queries, whereas dose tokens provide keys and values, allowing anatomical representations to incorporate spatially corresponding dose information. Conversely, in the second stream, dose tokens serve as queries, while CT tokens provide keys and values, enabling dosimetric representations to incorporate anatomical context. Each stream consists of two cross-attention blocks, with each block incorporating multi-head attention, residual connections, layer normalization, and a feed-forward multilayer perceptron (MLP). The resulting refined CT and dose features are combined to form a fused latent representation of dimension $2N\times L$, where $N=64$ denotes the number of tokens per modality and $L=256$ denotes the latent embedding dimension. In addition, modality-specific global representations are obtained by averaging the refined CT and dose features separately and retained for the contrastive alignment objective.

The fused CT and dose features are subsequently processed by a latent DiT consisting of six Transformer blocks, each with eight attention heads. Each block incorporates adaptive layer normalization conditioned on sinusoidal timestep embeddings, enabling timestep-dependent modulation of feature processing.

\textbf{Self-supervised representation learning with cross-modal alignment.} The self-supervised representation learning framework jointly optimizes three objectives: diffusion-based denoising to learn robust fused representations, cross-modal contrastive alignment to enhance the consistency between CT and dose features, and modality-specific feature reconstruction to preserve complementary information from each modality. These objectives are optimized without using xerostomia labels. Specifically, latent DiT progressively introduces Gaussian noise into the fused CT-dose representations over $T=1000$ diffusion time steps. Given the fused representation $\mathbf{z}_0$, the forward diffusion process generates a noisy representation at timestep $t$ as $\mathbf{z}_t=\sqrt{\bar{\alpha}_t}\mathbf{z}_0+\sqrt{1-\bar{\alpha}_t}\boldsymbol{\epsilon}, \quad \boldsymbol{\epsilon}\sim\mathcal{N}(\mathbf{0},\mathbf{I})$, where $\bar{\alpha}_t$ denotes the cumulative product of the noise-scheduling coefficients. The latent DiT is trained to predict the injected noise by minimizing the denoising objective as
\begin{align}
    \mathcal{L}_{\mathrm{diff}}=\mathbb{E}_{\mathbf{z}_0,t,\boldsymbol{\epsilon}}\left[\left|\boldsymbol{\epsilon}-\boldsymbol{\epsilon}_{\theta}(\mathbf{z}_t,t)\right|_2^2\right],
\end{align}
where $\boldsymbol{\epsilon}_{\theta}$ represents the DiT's predicted noise. This objective encourages the model to learn informative joint representations of CT anatomy and spatial dose distributions for downstream xerostomia risk prediction.

In addition to diffusion denoising, BA-DiT incorporates feature reconstruction and contrastive alignment objectives to preserve modality-specific information and promote cross-modal consistency between CT and dose representations, respectively. For feature reconstruction, the fused latent features are averaged along the dimension to obtain a global fused representation $\bar{\mathbf{z}}_0$. Two independent MLP decoders, $g_{\mathrm{CT}}$ and $g_{\mathrm{Dose}}$, map this representation to the reconstructed CT and dose features, $\hat{\mathbf{f}}_{\mathrm{CT}}$ and $\hat{\mathbf{f}}_{\mathrm{Dose}}$ by $\hat{\mathbf{f}}_{\mathrm{CT}}=g_{\mathrm{CT}}(\bar{\mathbf{z}}_0)$ and $\hat{\mathbf{f}}_{\mathrm{Dose}}=g_{\mathrm{Dose}}(\bar{\mathbf{z}}_0)$, respectively. The reconstruction targets, $\mathbf{f}_{\mathrm{CT}}$ and $\mathbf{f}_{\mathrm{Dose}}$, are obtained by global average pooling of the features from corresponding modality-specific encoders. Therefore, the feature reconstruction loss is defined by calculating the sum of the squared errors between the reconstructed features and their respective target features:
\begin{align}
    \mathcal{L}_{\mathrm{Recon}}=\left||\hat{\mathbf{f}}_{\mathrm{CT}}-\operatorname{sg}(\mathbf{f}_{\mathrm{CT}})|\right|_2^2+\left||\hat{\mathbf{f}}_{\mathrm{Dose}}-\operatorname{sg}(\mathbf{f}_{\mathrm{Dose}})|\right|_2^2,
\end{align}
where $\operatorname{sg}(\cdot)$ denotes the stop-gradient operation which is implemented to prevent gradients from propagating through the target features by detaching them from the computational graph. This objective encourages the fused representation to retain information relevant to both CT anatomy and spatial dose distributions.

For cross-modal alignment, the modality-specific global representations obtained after bidirectional cross-attention are independently projected through MLP-based projection heads and normalized to unit length. Given a minibatch of multiple paired CT and dose samples, the similarity between projected CT representation and projected dose representation is computed using cosine similarity scaled by a learnable temperature parameter. Symmetric InfoNCE contrastive learning is then applied to align representations belonging to the same patient while distinguishing them from mismatched CT–dose pairs within the minibatch. The contrastive objective is defined as
\begin{align}
    \mathcal{L}_{\mathrm{Contra}}=\frac{1}{2}\left(\mathcal{L}_{\mathrm{CT}\rightarrow\mathrm{dose}}+\mathcal{L}_{\mathrm{dose}\rightarrow\mathrm{CT}}\right),
\end{align}
where the two directional terms correspond to CT-to-dose and dose-to-CT contrastive classification, respectively. This symmetric objective promotes cross-modal consistency while retaining the distinct feature characteristics of the two modalities.

The overall self-supervised pretraining objective combines these three complementary losses as
\begin{align}
    \mathcal{L}_{\mathrm{SSL}}=\lambda_{\mathrm{Diff}}\mathcal{L}_{\mathrm{Diff}}+\lambda_{\mathrm{Recon}}\mathcal{L}_{\mathrm{Recon}}+\lambda_{\mathrm{Contra}}\mathcal{L}_{\mathrm{Contra}},    
\end{align}
where $\lambda_{\mathrm{diff}}$, $\lambda_{\mathrm{recon}}$, and $\lambda_{\mathrm{con}}$ are weighting coefficients controlling the contributions of diffusion denoising, feature reconstruction, and contrastive alignment, respectively. These objectives encourage the model to learn structured multimodal representations that capture cross-modal dependencies, preserve modality-specific information, and align paired CT and dose features without relying on toxicity labels.

\textbf{Supervised Bayesian fine-tuning for xerostomia risk prediction.} During supervised fine-tuning, the pretrained BA-DiT components are used to extract multimodal representations for xerostomia prediction. Specifically, the fused CT and dose features are passed through the pretrained DiT at a fixed feature-extraction timestep without adding diffusion noise. The resulting hidden representations are then used as inputs to the downstream classification module.  

To aggregate the DiT hidden features into a compact representation while preserving spatially informative features, a learnable attention-based spatial pooling module is employed. Specifically, the hidden features are first normalized using layer normalization and linearly projected from a 256-dimensional latent space into a 64-dimensional attention space. Four learnable query vectors are then applied to the projected token sequence through multi-head attention with four attention heads. The resulting four 64-dimensional query representations are concatenated to form a 256-dimensional pooled feature vector for classification.

The pooled feature vector is subsequently passed to a Bayesian classifier comprising two Bayesian linear layers with a hidden dimension of 128 and a ReLU activation. The weights and biases of each layer are modeled using mean-field Gaussian variational distributions, parameterized by trainable means $\mu$ and unconstrained parameters $\rho$, with standard deviations computed as $\sigma=\operatorname{softplus}(\rho)$. A zero-mean Gaussian  prior with a standard deviation of $1.0$ is assigned to the weights and biases. During inference, stochastic predictions are generated by sampling from the learned approximate posterior distributions using the local reparameterization trick. The resulting predictive distribution captures parameter uncertainty in the Bayesian classification head, conditional on the multimodal representations extracted by the pretrained feature-extraction modules. The final layer produces a scalar logit, which is transformed using the sigmoid function to obtain the predicted xerostomia probability, $p_{\mathrm{pred}}$.

To account for potential uncertainty in the observed xerostomia labels, a learnable label noise transition layer is incorporated into the framework. In this layer, two transition parameters, $\alpha$ and $\beta$, are optimized during supervised fine-tuning, and then transformed using sigmoid functions to constrain their values to the interval $(0, 1)$. Specifically, $\alpha$ represents the probability that a positive latent outcome is observed as negative, whereas $\beta$ represents the probability that a negative latent outcome is observed as positive. Given the predicted probability of the latent positive outcome $p_{\mathrm{pred}}$, the probability of observing a positive label $p_{\mathrm{obs}}$ is calculated as
\begin{align}
    p_{\mathrm{obs}}=p_{\mathrm{pred}}(1-\alpha)+(1-p_{\mathrm{pred}})\beta.
\end{align}
Therefore, the corresponding observed label logit is calculated from the observed xerostomia probability as
\begin{align}
    \ell_{\text{obs}} =\log\left(\frac{p_{\mathrm{obs}}}{1-p_{\mathrm{obs}}}\right).
\end{align}
This label noise transition layer distinguishes the predicted probability of the latent true outcome from that of the observed label, allowing the supervised optimization to account for potential label errors. When label-noise modeling is enabled, the observed label logit is used for computing the supervised loss, while the latent predicted probability is retained as the model's estimate of the underlying xerostomia risk.

The Bayesian classifier, spatial pooling module, and the label-noise model are jointly optimized using a binary cross entropy (BCE) loss $\mathcal{L}_{\mathrm{BCE}}$ computed from the observed xerostomia label logits. Additionally, the Kullback–Leibler (KL) divergence between the approximate posterior distributions and the corresponding Gaussian priors is computed for each Bayesian linear layer and incorporated into the objective as a variational regularization term $\mathcal{L}_{\mathrm{KL}}$. The overall supervised fine-tuning objective is therefore defined as the combination of the BCE loss and KL divergence regularization:
\begin{align}
   \mathcal{L}_{\mathrm{SF}}=\mathcal{L}_{\mathrm{BCE}}+\lambda_{\mathrm{KL}}\mathcal{L}_{\mathrm{KL}}. 
\end{align}
where $\lambda_{\mathrm{KL}}$ is set to $1\mathrm{e}{-2}$ to balance two components.

All experiments were implemented in PyTorch and conducted on an NVIDIA RTX A6000 GPU. A batch size of 4 was used for both self-supervised pre-training and supervised fine-tuning. During self-supervised pre-training, the model was trained for 500 epochs using the AdamW optimizer with an initial learning rate of $1\times10^{-4}$, and the learning rate was gradually reduced to $1\times10^{-5}$. For supervised Bayesian fine-tuning, the AdamW optimizer was employed, and the model was trained with an initial learning rate of $1\times10^{-4}$ and decayed with a cosine annealing scheduler.

\subsection{Risk Prediction and Uncertainty Quantification}
During Bayesian posterior sampling, $T$ stochastic forward passes ($T=200$) are performed, and the final xerostomia risk probability $\hat{p}$ is estimated by calculating mean of the true prediction probability distribution:
\begin{align}
	\hat{p} = \frac{1}{T} \sum_{t=1}^{T} p_{\mathrm{pred}}^{(t)}.
\end{align}
where $p_{\mathrm{pred}}^{(t)}$ denotes the predicted probability from the t-th posterior sample. To derive a discrete classification for xerostomia grades, a decision threshold $\tau$ is applied to the xerostomia risk probability $\hat{p}$, converting it into a binary label $\hat{y}$ via the indicator function as
\begin{align}
	\hat{y} = \mathds{1}[\hat{p} \geq \tau].
\end{align}
Three types of predictive uncertainties are quantified from the sampled probability distribution: aleatoric uncertainty ($U_{\text{alea}}$), epistemic uncertainty ($U_{\text{epst}}$), and label uncertainty ($U_{\text{label}}$), capturing uncertainty arising from inherent outcome variability, limited model knowledge, and potential inaccuracies in the observed toxicity labels, respectively.

\textbf{Aleatoric uncertainty} ($U_{alea}$) captures inherent stochasticity of the input data and its impact on model predictions. For the binary classification task, it is quantified by averaging the variances of the probability distribution across $T$ stochastic forward passes generated by sampling from the approximate posterior distributions of the Bayesian classifier \cite{nair2020exploring,herzog2020integrating}:
\begin{align}
	U_{alea}=\frac{1}{T} \sum_{t=1}^{T} p_{\mathrm{pred}}^{(t)} \big(1 - p_{\mathrm{pred}}^{(t)}\big).
\end{align}
This metric effectively identifies cases where the underlying data lacks sufficient discriminative features to produce a certain prediction.

\textbf{Epistemic uncertainty} ($U_{epst}$), also known as model uncertainty, reflects limitations in the model's knowledge due to insufficient or non-representative training data. In clinical settings, high $U_{\text{epst}}$ may indicate that a patient exhibits characteristics, such as unusual anatomical features or uncommon patterns, that are underrepresented in the training data distribution. Such cases may warrant additional clinical review or data acquisition. Epistemic uncertainty is estimated by calculating the standard deviation of the predicted probabilities across $T$ stochastic forward passes \cite{herzog2020integrating}: 
\begin{align}
	U_{\text{epst}} = \sqrt{\frac{1}{T} \sum_{t=1}^{T} \left(p_{\mathrm{pred}}^{(t)} - \hat{p}\right)^2}.
\end{align}
Unlike aleatoric uncertainty, epistemic uncertainty is potentially reducible through additional training data and improved model representation.

\textbf{Label uncertainty} ($U_{\text{label}}$) quantifies the uncertainty associated with potential errors in the observed toxicity annotations, as modeled by the learnable label noise transition process described above. Given the xerostomia risk probability ($\hat{p}$) and the learned transition parameters ($\alpha$ and $\beta$), label uncertainty is quantified as the expected variance associated with the label transitions \cite{kendall2017uncertainties}:
\begin{align}
	U_{\text{label}} =\hat{p} \cdot \alpha (1-\alpha) + (1 - \hat{p}) \cdot \beta (1-\beta).
\end{align}
It characterizes the contribution of potential label errors to predictive uncertainty, complementing the aleatoric and epistemic uncertainty estimates.

\subsection{Model Evaluation}
The model was evaluated in terms of predictive performance, probability calibration, and clinical utility, as described below.

\noindent\textbf{Predictive performance.} Model performance was evaluated using five classification metrics: area under the receiver operating characteristic curve (AUC), accuracy, sensitivity, specificity, and F1 score. Given the class imbalance across cohorts, AUC was designated as the primary metric because it provides a threshold-independent assessment of discriminative performance. The remaining metrics were reported to characterize classification performance at the prespecified decision threshold. Five-fold cross validation was implemented for evaluation and independently repeated using two random initialization seeds (0 and 42) to assess reproducibility. Paired DeLong tests were used to compare the AUCs of BA-DiT and competing models, with false discovery rate (FDR) correction applied for multiple comparisons.

\noindent\textbf{Probability calibration.} Model calibration was evaluated using the expected calibration error (ECE) \cite{pakdaman2015}, which quantifies the discrepancy between predicted xerostomia probabilities and outcome frequencies across probability bins. Lower ECE values indicate better calibration, with an ECE of zero representing perfect agreement between predicted probabilities and observed event frequencies.

\noindent\textbf{Clinical utility.} Decision curve analysis (DCA) \cite {vickers2006, vickers2021decision} was performed to evaluate the potential clinical utility of BA-DiT across a range of risk threshold probabilities. To further investigate the contribution of uncertainty estimation to clinical utility, DCA was conducted both in the full cohort and in sub-cohorts using an uncertainty-based selective prediction strategy. Specifically, two uncertainty thresholds were defined based on the empirical distribution of total predictive uncertainty. For each threshold, patients with uncertainty exceeding the threshold were excluded from the selective prediction analysis, thus evaluating the clinical utility of model predictions among patients with lower predictive uncertainty at different levels of prediction coverage. Net benefit was evaluated across a range of threshold probabilities and compared among the conventional BA-DiT model, the two uncertainty-aware selective prediction strategies, and the standard treat-all and treat-none strategies. This analysis assessed whether withholding predictions for patients with higher predictive uncertainty improved the net benefit of BA-DiT across clinically relevant thresholds.

\noindent\textbf{Model interpretation.} To further examine the predictive behavior of BA-DiT, representative patient cases from the MSHS and Emory cohorts were selected. Three-dimensional gradient-weighted class activation mapping (3D Grad-CAM) \cite{Selvaraju2019} was employed to visualize spatial regions contributing to model predictions, providing insights into the anatomical and dosimetric features associated with predicted toxicity risks.

\section{Results}
\subsection{Performance in Risk Prediction}
To evaluate the effectiveness of BA-DiT, we compared its performance with several baseline and ablation models to examine the contributions of multimodal integration, bidirectional cross-attention fusion, and the DiT architecture. Specifically, BA-DiT was compared with unimodal 3D CNN and 3D DiT models using either pretreatment CT images (CT Only) or planning dose distributions (Dose Only) to highlight the importance of using anatomical and dosimetric information as multimodal inputs. We further compared BA-DiT with a multimodal 3D CNN and 3D DiT using channel-wise concatenation to assess the superiority of bidirectional cross-attention fusion module. Finally, 3D CNN models integrated with bidirectional cross-attention fusion were evaluated to isolate the contributions of DiT on latent feature encoding.

For the first evaluation, data from three independent cohorts (Moffitt, Emory, and TCIA) were combined for pretraining and supervised fine-tuning of BA-DiT and all baseline models. The trained models were subsequently evaluated on the independent MSHS cohort (first half of Table \ref{tab1}). BA-DiT achieved the highest AUC, accuracy, sensitivity and F1 score, significantly outperforming all baseline models, with an ECE of $0.050$. Multimodal models generally outperformed the unimodal counterparts, while the 3D DiT models generally achieved better performance than their corresponding 3D CNN models. Incorporating bidirectional cross attention fusion also improved the performance of the CNN-based multimodal model compared with channel-wise concatenation.

In the second evaluation, BA-DiT and all baseline models were trained using data from the Moffitt, MSHS, and TCIA cohorts and evaluated on the independent Emory cohort (second half of Table \ref{tab1}). BA-DiT achieved the highest AUC, accuracy, sensitivity, and F1 score, while maintaining favorable specificity, significantly outperforming all baseline models, with an ECE of 0.089. Additionally, multimodal models generally outperformed their unimodal counterparts, and the 3D DiT models generally outperformed their 3D CNN models. Bidirectional cross attention fusion also improved the performance of the CNN-based multimodal model compared with channel-wise concatenation.

\begin{table*}[!t]
\centering
\caption{Performance comparison of BA-DiT and baseline models for xerostomia risk prediction in the independent MSHS and Emory cohorts. Model performance was assessed using the area under the receiver operating characteristic curve (AUC), accuracy, specificity, sensitivity, and F1 score. Results are presented as mean $\pm$ standard deviation (SD). \textbf{Bold} indicates the best performance for each metric. AUC comparison between BA-DiT and each baseline model were performed using statistical tests, with p-values adjusted for multiple comparisons using the false discovery rate (FDR) procedure.}
\label{tab1}
\resizebox{\textwidth}{!}{
\begin{tabular}{c|c|c|ccccc|c}
\toprule
Cohorts & Models & Multimodal Fusion & AUC & Accuracy & Specificity & Sensitivity & F1 Score & $p_{\textrm{adj}}$ \\
\midrule
\multirow{8}{*}{\rotatebox[origin=c]{90}{MSHS}}
& 3D CNN & CT Only   & 49.27$\pm$2.38 & 37.30$\pm$5.61 & 88.00$\pm$16.97 & 3.95$\pm$1.86  & 6.92$\pm$2.52 & $<0.001$ \\
& 3D CNN & Dose Only & 55.69$\pm$0.45 & 52.38$\pm$0.00 & 48.00$\pm$5.66  & 55.26$\pm$3.72 & 58.30$\pm$1.64 & $<0.001$ \\
& 3D DiT & CT Only   & 61.95$\pm$0.08 & 41.27$\pm$2.25 & \textbf{100.00}$\pm$0.00 & 2.63$\pm$3.72  & 5.00$\pm$7.07 & $<0.001$ \\
& 3D DiT & Dose Only & 68.53$\pm$0.59 & 67.46$\pm$1.12 & 50.00$\pm$2.83  & 78.95$\pm$0.00  & 74.54$\pm$0.66 & $<0.001$ \\
& 3D CNN & CT-Dose Concatenation & 65.47$\pm$2.23 & 52.38$\pm$6.73 & 74.00$\pm$2.83 & 38.16$\pm$3.03 & 48.57$\pm$2.12 & $<0.001$ \\
& 3D CNN & CT-Dose Bi-attention & 66.63$\pm$3.88 & 58.73$\pm$2.25 & 74.00$\pm$2.83 & 48.69$\pm$1.86 & 58.73$\pm$2.25 & $<0.001$ \\
& 3D DiT & CT-Dose Concatenation & 72.27$\pm$1.56 & 71.43$\pm$4.49 & 60.00$\pm$5.66 & 78.95$\pm$3.72 &  76.93$\pm$3.63 & $<0.001$ \\
& BA-DiT & CT-Dose Bi-attention & \textbf{83.11}$\pm$3.05 & \textbf{81.75}$\pm$1.12 & 74.00$\pm$2.83 & \textbf{86.84}$\pm$3.72 & \textbf{85.15}$\pm$1.32 & - \\
\midrule
\multirow{8}{*}{\rotatebox[origin=c]{90}{Emory}}
& 3D CNN & CT Only   & 52.78$\pm$6.84 & 34.01$\pm$0.00 & \textbf{100.00}$\pm$0.00 & 0.00$\pm$0.00 & 0.00$\pm$0.00 & $<0.001$ \\
& 3D CNN & Dose Only & 56.71$\pm$2.35 & 43.20$\pm$7.21 & 81.00$\pm$4.24 & 23.71$\pm$13.12 & 34.62$\pm$15.68 & $<0.001$ \\
& 3D DiT & CT Only   & 54.36$\pm$7.45 & 39.46$\pm$0.00 & 88.00$\pm$8.49 & 14.44$\pm$4.38 & 23.73$\pm$5.53 & $<0.001$ \\
& 3D DiT & Dose Only & 60.22$\pm$0.77 & 45.58$\pm$4.81 & 80.00$\pm$8.49 & 27.84$\pm$11.66 & 39.59$\pm$12.30 & $<0.001$ \\
& 3D CNN & CT-Dose Concatenation & 60.73$\pm$0.89 & 54.08$\pm$0.48 & 73.00$\pm$4.24 & 44.33$\pm$2.91 & 56.00$\pm$1.87 & $<0.001$ \\
& 3D CNN & CT-Dose Bi-attention  & 64.90$\pm$2.64 & 57.14$\pm$2.89 & 73.00$\pm$21.21 & 48.97$\pm$15.39 & 59.39$\pm$9.23 & $<0.001$ \\
& 3D DiT & CT-Dose Concatenation & 66.11$\pm$1.96 & 64.63$\pm$1.92 & 54.00$\pm$0.00 & 70.10$\pm$2.91 & 72.33$\pm$1.92 & $<0.001$ \\
& BA-DiT               & CT-Dose Bi-attention & \textbf{80.33}$\pm$0.00 & \textbf{77.21}$\pm$0.48 & 77.00$\pm$1.41 & \textbf{77.32}$\pm$0.00 & \textbf{81.75}$\pm$0.32 & - \\
\bottomrule
\end{tabular}}
\end{table*}

\subsection{Uncertainty and Impact on Clinical Utility}
\begin{figure*}[!t]
    \centering
    \begin{subfigure}[t]{0.49\linewidth}
        \centering
        \includegraphics[width=\linewidth]{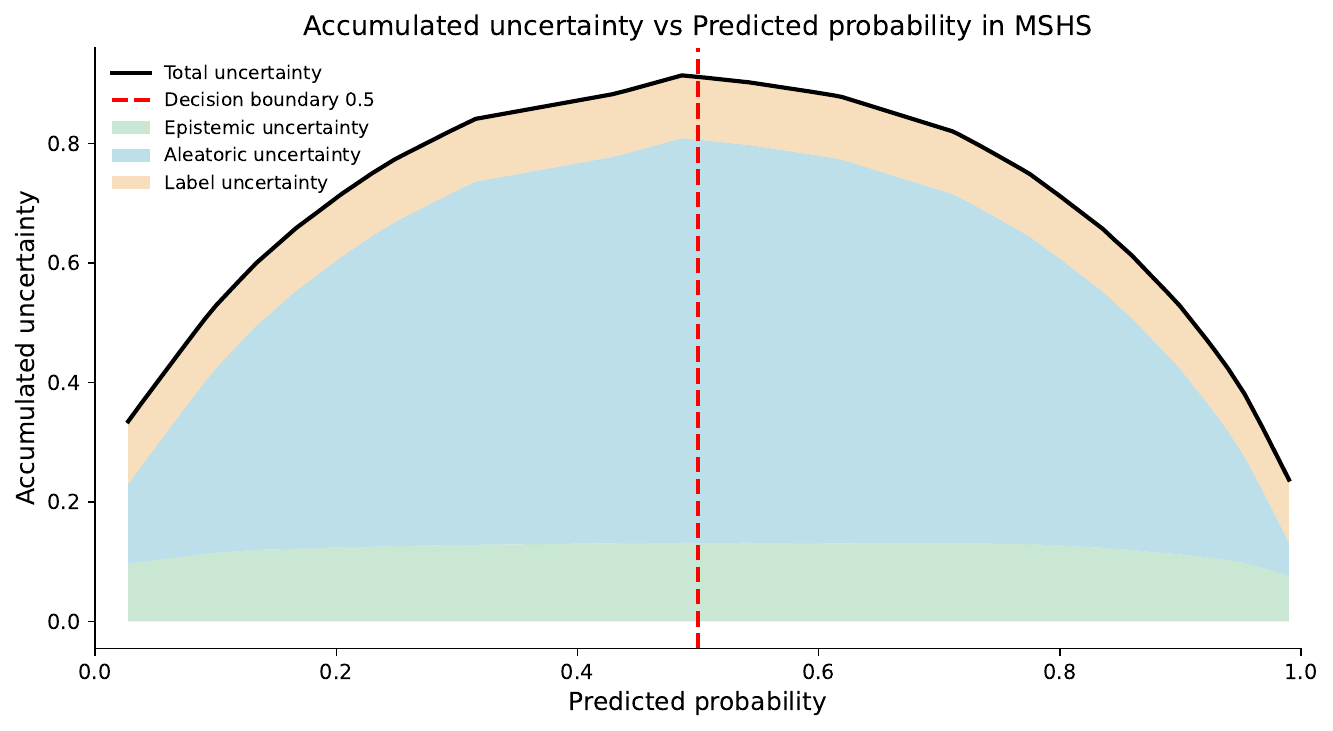}
        \label{fig:uncertainty_mshs}
    \end{subfigure}
    \hfill
    \begin{subfigure}[t]{0.49\linewidth}
        \centering
        \includegraphics[width=\linewidth]{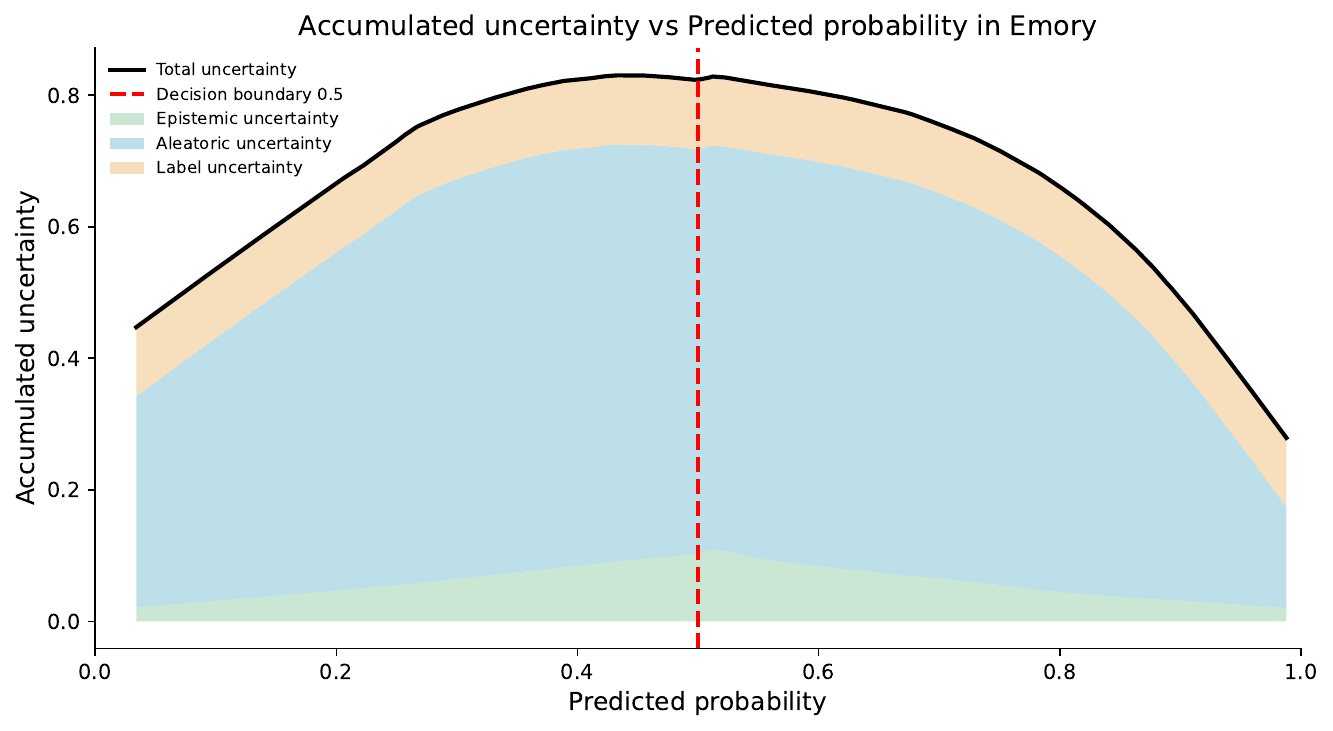}
        \label{fig:uncertainty_emory}
    \end{subfigure}
    \caption{Predictive uncertainty as a function of predicted xerostomia probability across the MSHS and Emory cohorts. Total predictive uncertainty is decomposed into aleatoric, epistemic, and label uncertainty. Aleatoric and epistemic uncertainty increase near the decision boundary (predicted probability $\approx 0.5$) and decrease toward the probability extremes, whereas label uncertainty shows a less pronounced increase near the decision boundary.}
    \label{fig:uncertainty}
\end{figure*}

Figure~\ref{fig:uncertainty} illustrates the distributions of total predictive uncertainty and its aleatoric, epistemic, and label uncertainty across predicted xerostomia probabilities in the MSHS and Emory cohorts. Aleatoric uncertainty was the predominant component of total predictive uncertainty in both cohorts. Total uncertainty was highest for predictions near the decision boundary (predicted probability of approximately 0.5) and decreased toward the extremes of the probability range. Epistemic uncertainty showed a similar pattern, with higher values near the decision boundary and lower values for predictions closer to 0 or 1. In contrast, label uncertainty exhibited less variation across the predicted probability range. Overall, predictions with intermediate estimated xerostomia probability were associated with greater total uncertainty with aleatoric and epistemic uncertainty contributing most prominently to this pattern.

\begin{figure*}[!t]
    \centering
    \begin{subfigure}[t]{0.49\linewidth}
        \centering
        \includegraphics[width=\linewidth]{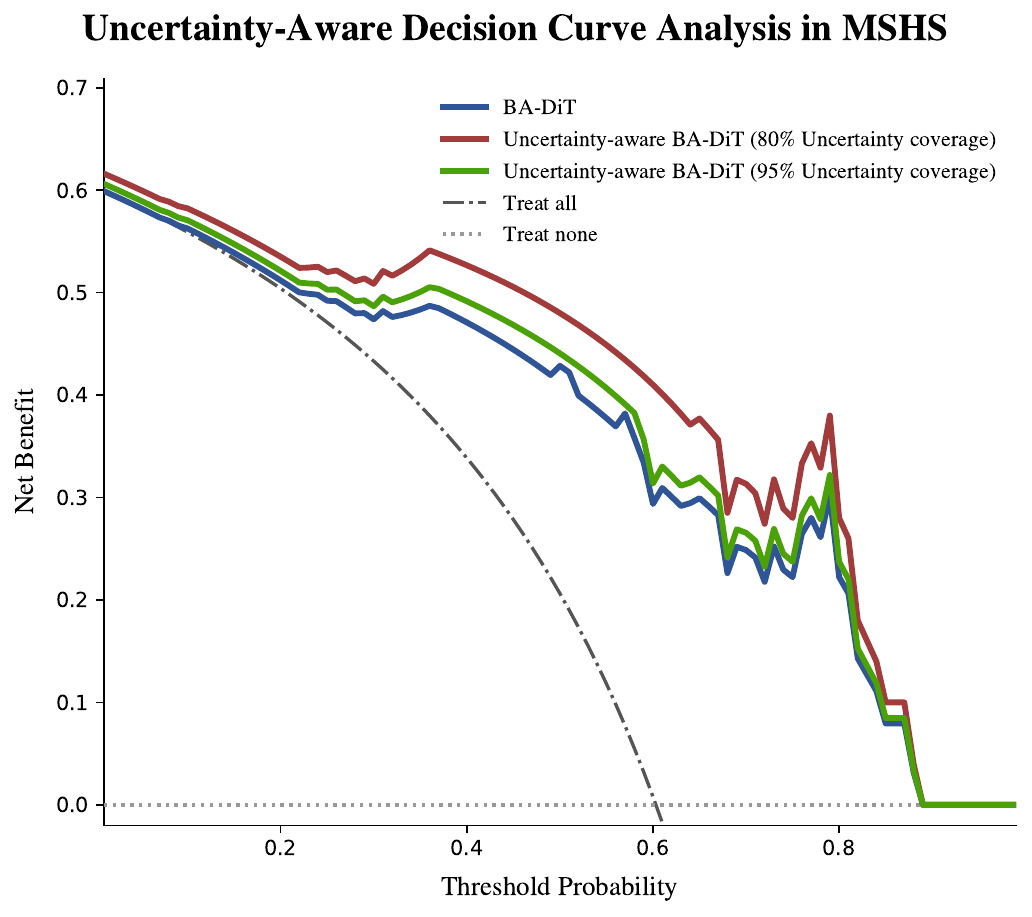}
    \end{subfigure}
    \hfill
    \begin{subfigure}[t]{0.49\linewidth}
        \centering
        \includegraphics[width=\linewidth]{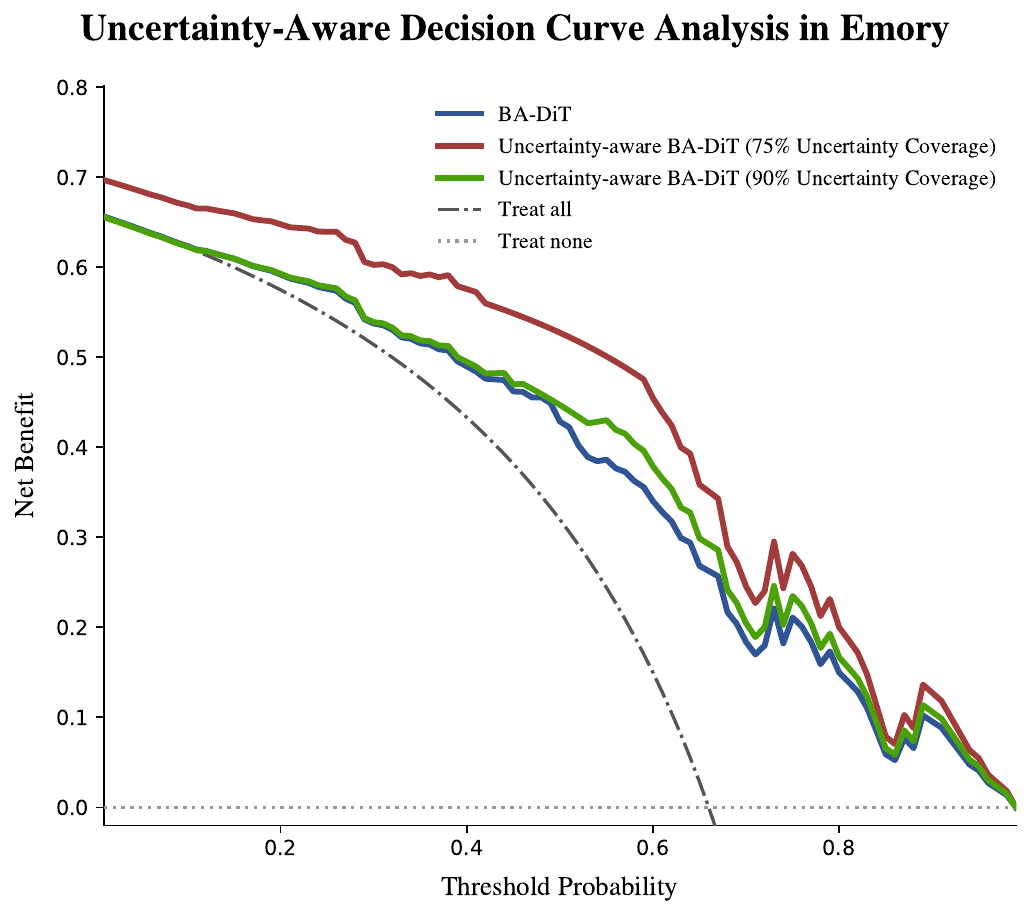}
    \end{subfigure}
    \caption{Uncertainty-aware decision curve analysis for radiation xerostomia risk prediction in the MSHS and Emory cohorts. Net benefit is shown across a range of threshold probabilities for the conventional BA-DiT model and two uncertainty-aware BA-DiT strategies using the 80th- and 95th-percentile and 75th- and 90th-percentile total-uncertainty thresholds in the two cohorts, respectively. Patients with uncertainty above each threshold were excluded from the corresponding selective prediction analysis. “Treat all” and “Treat none” are included as reference decision strategies.}
    \label{fig:decision}
\end{figure*}

DCA was performed to evaluate the potential clinical utility of BA-DiT by evaluating its net benefit across a range of threshold probabilities and comparing with the standard treat-all and treat-none strategies in the independent MSHS and Emory cohorts (Figure \ref{fig:decision}). We further evaluated the importance of individualized uncertainty quantification on improving clinical utility by evaluating the net benefit of uncertainty-based selective prediction. Specifically, in addition to evaluating BA-DiT in the full cohort, we performed selective DCA after excluding patients with higher predictive uncertainty. In the MSHS cohort, the two uncertainty thresholds corresponded to the 80th and 95th percentiles of the empirical distribution of total predictive uncertainty, resulting in the exclusion of approximately the highest-uncertainty $20\%$ and $5\%$ of patients, respectively. In the Emory cohort, the thresholds corresponded to the 75th and 90th percentiles, excluding approximately the highest-uncertainty $25\%$ and $10\%$ of patients, respectively. In both cohorts, BA-DiT achieved positive net benefit over a broad range of threshold probabilities and generally outperformed the treat-all and treat-none strategies. In the MSHS cohort, both uncertainty-aware selective prediction strategies yielded higher net benefit than the conventional BA-DiT model across much of the evaluated threshold range, with a more pronounced improvement for the $80\%$ prediction-coverage strategy (excluding approximately $20\%$ of patients with the highest uncertainty). These findings indicate that excluding patients with higher predictive uncertainty can improve the net benefit of model-guided decision-making among the remaining patients. Similar patterns were observed in the Emory cohort.

\subsection{Illustrative Cases of Uncertainty-Aware Risk Prediction}
Four representative patient cases from the MSHS and Emory cohorts were selected to illustrate two complementary prediction scenarios (Table~\ref{tab2}): (1) correctly classified positive cases with high predicted xerostomia risk and relatively low predictive uncertainty, and (2) negative cases with predictions near the decision boundary and substantially higher predictive uncertainty.

\begin{figure*}[t]
	\centering
	\includegraphics[width=\linewidth]{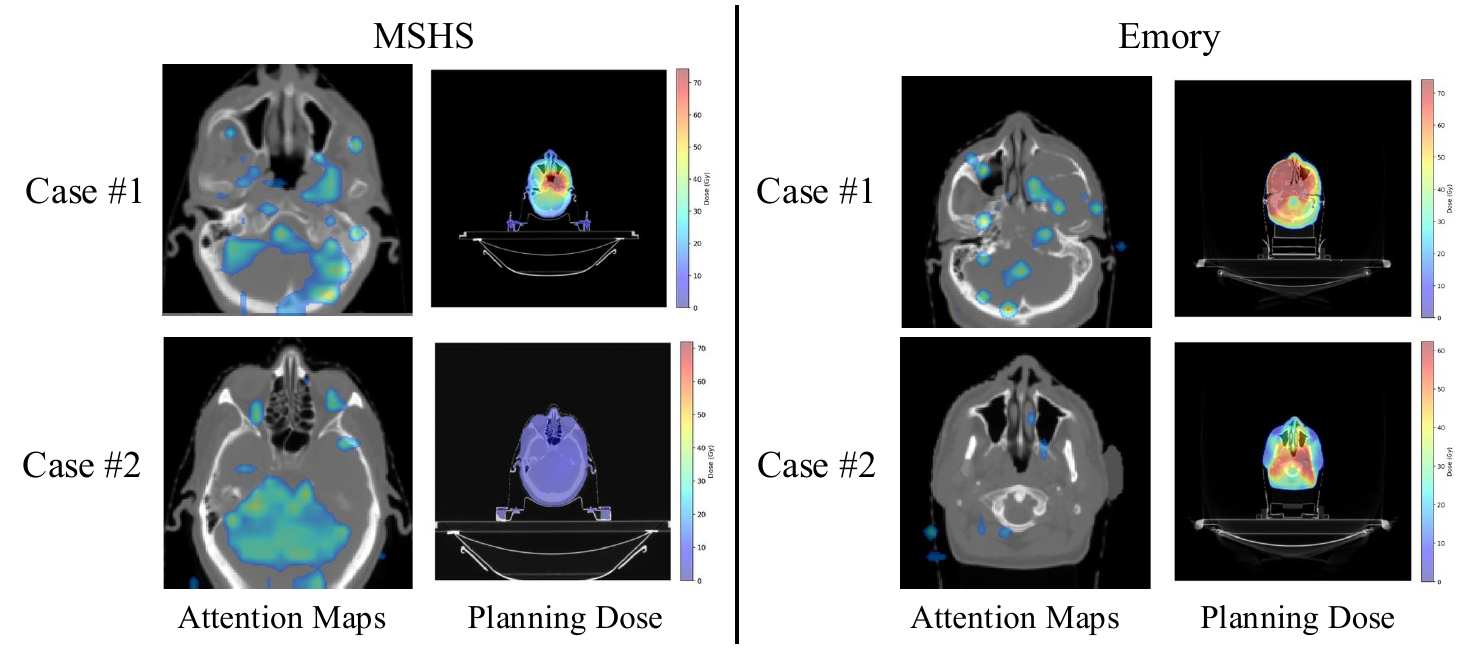} \\
	\caption{Visualization of attention maps in anatomical regions contributing to radiation xerostomia predictions and their spatial relationship with radiation dose. Two representative cases (Cases \#1 and \#2) from the MSHS and Emory cohorts are shown, respectively. The left rows in each cohort show attention maps on pretreatment CT images via Grad-CAM, illustrating the model attribution in risk prediction and uncertainty quantification, while the right rows show the corresponding planning CT images with radiation dose distributions. Warmer colors in the Grad-CAM maps indicate greater relative model attribution, whereas the dose overlays represent the planning radiation dose (Gy).}
	\label{fig:interpretation}
\end{figure*}

\noindent\textbf{Case 1: Correctly classified positive predictions with relatively low uncertainty.} BA-DiT predicted high xerostomia probabilities of $0.928$ and $0.959$ for the MSHS and Emory cohorts, respectively, correctly classifying both patients as positive. Both predictions were associated with relatively low predictive uncertainty, particularly in the MSHS case. The corresponding 3D Grad-CAM maps showed prominent activation in regions largely overlapping with target anatomical structures receiving relatively high radiation doses (Figure~\ref{fig:interpretation}). These visualizations suggest that BA-DiT confidently captured correct spatial information from the anatomical structures and their associated dose distributions contributed to the model's positive predictions.

\noindent\textbf{Case 2: Near-boundary predictions with high predictive uncertainty.} BA-DiT predicted xerostomia probabilities of $0.508$ and $0.533$ for the MSHS and Emory cohorts, respectively. Both predictions were close to the decision boundary ($\tau=0.5$) and resulted in false-positive classifications despite negative ground-truth labels. Compared with Case 1 within the corresponding cohorts, these cases exhibited substantially higher epistemic and aleatoric uncertainty. Specifically, epistemic uncertainty increased approximately 12-fold and 2-fold, and aleatoric uncertainty increased approximately 2.7-fold and 2.9-fold in the MSHS and Emory cohorts, respectively. The corresponding 3D Grad-CAM visualizations provided additional insight into the spatial patterns associated with these predictions. In the MSHS case, the highlighted regions showed limited overlap with target structures receiving relatively high radiation doses, with greater activation in regions which were irrelevant to xerostomia and exposed to lower doses. A similar pattern was observed in the Emory case, where the highlighted regions showed limited correspondence with target structures receiving high-dose exposure. These visual patterns suggest that the model's predictions in these cases were less strongly associated with the expected high-dose anatomical regions, so the model made incorrectly risk predictions with high uncertainty.

\begin{table*}[!t]
\centering
\caption{Four representative patient cases illustrating uncertainty-aware xerostomia prediction in the MSHS and Emory cohorts. Predicted xerostomia probabilities and their corresponding uncertainty components, including epistemic, aleatoric, and label uncertainty, are reported for each case. Cases 1 and 2 from each cohort represent correctly classified positive predictions with relatively low uncertainty and false-positive predictions near the decision boundary with high uncertainty, respectively.}
\label{tab2}
\resizebox{\textwidth}{!}{
\begin{tabular}{c|c|cc|ccc}
\toprule
Cohorts & Case & Label & Xerostomia risk & $U_{\mathrm{epst}}$ & $U_{\mathrm{alea}}$ & $U_{\mathrm{label}}$\\
\midrule
\multirow{2}{*}{MSHS}
& 1 & 1 & 0.928 & 0.010 & 0.248 & 0.105 \\
& 2 & 0 & 0.508 & 0.123 & 0.672 & 0.105 \\
\midrule
\multirow{2}{*}{Emory}
& 1 & 1 & 0.959 & 0.156 & 0.215 & 0.105 \\
& 2 & 0 & 0.533 & 0.318 & 0.618 & 0.105\\
\bottomrule
\end{tabular}}
\end{table*}

\section{Discussion}\label{sec4}
BA-DiT demonstrated superior predictive performance compared with unimodal models and alternative multimodal approaches in the independent MSHS and Emory cohorts, achieving AUCs of $0.831\%$ and $0.803\%$, respectively, along with the highest accuracy, sensitivity, and F1 scores among the evaluated models (Table \ref{tab1}). The consistent results across these cohorts support the potential superiority and generalizability of the framework. Multimodal models generally outperformed their unimodal counterparts, supporting the value of integrating pretreatment CT images with RT dose distributions for xerostomia prediction. Notably, dose-only models consistently outperformed corresponding CT-only models, highlighting the predictive relevance of spatial dose information. Nevertheless, the additional gains from multimodal integration suggest that CT-derived anatomical features provide complementary information beyond dose alone. Multimodal models incorporating bidirectional cross attention fusion also generally outperformed those using channel-wise concatenation, indicating that explicitly modeling cross-modal interactions may improve the integration of anatomical and dosimetric information. Finally, the improved performance of 3D DiT models over corresponding CNN counterparts suggests that Transformer-based representations may better capture complex spatial relationships in these data.

Beyond predictive performance, uncertainty quantification provided complementary information for interpreting individual xerostomia risk estimates. Aleatoric uncertainty was the predominant component and was generally highest for predictions near the decision boundary, consistent with the greater uncertainty associated with probabilities near 0.5 (Figure \ref{fig:uncertainty}). This measure should not, however, be interpreted exclusively as intrinsic biological variability or as evidence of inadequate anatomical or dosimetric information. It may also reflect unexplained variability related to unmeasured clinical factors, differences in baseline salivary function, and heterogeneous post-RT responses. Xerostomia is a multifactorial toxicity influenced by radiation exposure, baseline salivary function, patient characteristics, and treatment-related factors \cite{dirix2006radiation,vissink2010clinical,beetz2012ntcp,beetz2014quantec}. Thus, patients with similar pretreatment anatomical and dosimetric characteristics may experience different degrees of salivary dysfunction. Because the current framework uses only pretreatment CT and planned dose distributions, some of this variability may remain unexplained. Incorporating baseline salivary function, patient-reported symptoms, and longitudinal clinical assessments, potentially complemented by functional imaging, may improve the characterization of individual risk in future studies. 

Epistemic uncertainty reflects uncertainty associated with the model's learned parameters and may increase when patient characteristics, imaging patterns, or treatment profiles are insufficiently represented in the training data. This is particularly relevant in across-institutional applications, where differences in patient populations, imaging protocols, and clinical workflows may introduce distribution shifts \cite{wahid2024artificial,dunger2025uncertainties,lambert2024trustworthy}. Such estimates may help identify predictions for which the model has limited support from its learned representations and that therefore warrant more cautious interpretation or further evaluation. In the current framework, this estimate primarily reflects parameter uncertainty in the Bayesian classification component, conditional on the extracted multimodal representations.

Label uncertainty accounts for potential inconsistencies in the observed xerostomia outcomes. Clinical assessment may vary with grading criteria, assessment timing, documentation practices, and the use of clinician-reported versus patient-reported measures \cite{dirix2006radiation,vissink2010clinical,beetz2012ntcp}. The learnable label-transition model was incorporated to account for potential errors in observed toxicity labels during model training. However, it cannot fully resolve uncertainty arising from incomplete documentation or heterogeneous outcome assessment, nor can it identify the specific source of ambiguity in an individual patient's outcome. Together, these three uncertainty estimates provide information beyond the predicted xerostomia probability and may help identify cases requiring more cautious interpretation.

The uncertainty-aware DCA further explored the potential clinical relevance of combining predicted xerostomia risk with predictive uncertainty (Figure \ref{fig:decision}). Uncertainty-based selective prediction generally improved net benefit among patients retained for prediction, suggesting that accounting for predictive uncertainty may help inform the use of model-generated risk estimates \cite{vickers2021decision}. For example, patients with high predicted risk and relatively low uncertainty may warrant closer consideration for preventive measures or supportive care, whereas those with greater uncertainty may benefit from additional assessment of baseline salivary function, symptom burden, and other relevant clinical factors. This framework is not intended to independently determine treatment modifications or replace clinical assessment. Importantly, excluding patients with high predictive uncertainty changes the composition of the evaluated population and may alter its outcome prevalence. Therefore, improved net benefit among patients retained for prediction does not necessarily translate into greater benefit at the population level or improved clinical outcomes. Prospective studies are needed to determine whether uncertainty-guided risk assessment can improve xerostomia surveillance, facilitate timely supportive interventions, or inform individualized radiotherapy planning.

The illustrative patient cases further demonstrate how BA-DiT combines risk prediction with uncertainty estimation to distinguish predictions with different level of confidence. Cases with high predicted risk and relatively low uncertainty were contrasted with near-boundary predictions associated with substantially higher uncertainty, highlighting the potential value of uncertainty decomposition in identifying challenging cases. The accompanying 3D Grad-CAM visualizations (Figure~\ref{fig:interpretation}) provide complementary spatial insights into the image regions contributing to individual predictions. Differences in highlighted regions across these cases suggest that the spatial features contributing to model predictions may vary with patient risk profiles. Although these visualizations do not establish causal relationships or directly explain predictive uncertainty, they offer an additional perspective for examining model behavior and interpreting individualized predictions.

Several limitations should be considered. First, this was a retrospective study, and BA-DiT relied primarily on pretreatment CT images and RT dose distributions without incorporating clinical variables, such as demographic characteristics, baseline salivary function, or comorbidities. This was partly due to the limited availability and inconsistent documentation of these variables across the four cohorts. Although imaging and dosimetric features provide complementary information, they may not fully capture individual susceptibility to xerostomia. Future studies should investigate the integration of harmonized clinical variables and patient-reported outcomes to improve risk prediction. 

Second, although the model was developed using data from multiple cohorts and evaluated in independent MSHS and Emory cohorts, broader external validation is needed to establish its generalizability across diverse patient populations, imaging protocols, treatment techniques, and clinical settings \cite{mali2021making, barragan2022towards}. Differences in CT acquisition, dose calculation, and toxicity assessment may introduce distribution shifts that affect both predictive performance and uncertainty estimation. Future validation should therefore assess not only discrimination and calibration but also the reliability of uncertainty estimates across different clinical and technical settings. 

Third, the study focused on xerostomia assessed at a single time point, six months after RT treatment, although salivary dysfunction may evolve over time \cite{berger2022predicting, tzikas2023ntcp, hiraoka2024long}. This endpoint does not capture the onset, duration, progression, or long-term recovery of xerostomia. Prospective studies with standardized toxicity grading, baseline salivary function assessments, patient-reported outcomes, and longitudinal follow-up are needed to establish its temporal course more comprehensively. 

Finally, the thresholds used for risk-uncertainty stratification were selected for analytical purposes rather than derived from clinically optimized decision criteria. Their clinical applicability, therefore, remains to be established. Further studies are needed to determine clinically meaningful thresholds and evaluate whether this approach can support toxicity surveillance and supportive care.

\section{Conclusion}
In this study, we developed an uncertainty-aware multimodal framework, termed BA-DiT, for individualized prediction of radiation-induced xerostomia in patients with HNC. BA-DiT demonstrated superior predictive performance over baseline models in two independent cohorts while providing complementary estimates of predictive uncertainty. These findings highlight the potential value of combining anatomical and dosimetric information with uncertainty quantification for personalized xerostomia risk assessment.

\section*{Acknowledgment}
This work was partially supported by the Radiation Oncology Institute (Grant \#ROI 2022-9132) and the HESI THRIVE Initiative.

\section*{Conflict of interests}
The authors declare no conflicts of interest.

\bibliography{references}

\end{document}